\documentclass[letterpaper, 10 pt, conference]{IEEEtran}

\usepackage{amsfonts,amsmath,amsthm,amscd,amssymb,array}
\usepackage{MnSymbol}

\usepackage[T1]{fontenc}
\usepackage[latin1]{inputenc}

\usepackage{nicefrac}				
\usepackage{fancyhdr,xcolor,fancybox}
\usepackage{algpseudocode,algorithm}		
\usepackage{makeidx} 				
\usepackage{pstricks,pst-plot,pst-node} 	
\usepackage{url}				
\usepackage{graphicx}				
\usepackage{listings}				
\usepackage{subfigure}				
\usepackage{enumerate}				
\usepackage[intoc,english]{nomencl}		
\usepackage{acronym}				
\usepackage{cite}				
\usepackage{ifsym}
\usepackage{xspace}
\usepackage{ulem}
\usepackage{array}
\usepackage{tabularx}
\usepackage{flushend}				
\usepackage{marvosym}

\usepackage{etoolbox}				

\newtoggle{EXTENDED}				

\newtheorem{theorem}{Theorem}					
\newtheorem{lemma}[theorem]{Lemma}				

\newtheorem{corollary}[theorem]{Corollary}				

\newcommand{\eg}{\mbox{e.\,g.}\xspace}				
\newcommand{\ie}{\mbox{i.\,e.}\xspace}				
\newcommand{\wrt}{\mbox{w.\,r.\,t.}\xspace}			
\newcommand{\etal}{\mbox{et\,al.}\xspace}			

\newcommand{\ts}{dimension\xspace}

\newcommand{\Tx}{\mathrm{Tx}}					
\newcommand{\Rx}{\mathrm{Rx}}					
\renewcommand{\mod}{\mathrm{mod}}				
\newcommand{\DoF}{\mathrm{DoF}}					
\newcommand{\modxn}{\ \mod\,(x^n-1)}				

\IEEEoverridecommandlockouts

\toggletrue{EXTENDED}

\begin{document}
\addtolength{\voffset}{0.3cm}
\title{Cyclic Interference Alignment and Cancellation \\ in $3$-User $X$-\,Networks with Minimal Backhaul}

\author{
\IEEEauthorblockN{
Henning Maier 
and Rudolf Mathar 
}

\\

\IEEEauthorblockA{
Institute for Theoretical Information Technology\\
RWTH Aachen University,
52056 Aachen, Germany\\
Email: \{maier, mathar\}@ti.rwth-aachen.de}
\thanks{This work has been supported by the \emph{Deutsche Forschungs\-gemein\-schaft} (DFG) within the project \emph{Power
Adjustment and Constructive Interference Alignment for Wireless Networks} (PACIA - Ma 1184/15-2) of the DFG program
\emph{Communication in Interference Limited Networks} (COIN) and furthermore by the UMIC Research Centre, RWTH Aachen University.}
}

\maketitle
\IEEEpeerreviewmaketitle

\begin{abstract}
\iftoggle{EXTENDED}
{}
{\emph{To be considered for an IEEE Jack Keil Wolf ISIT Student Paper Award.}}
We consider the problem of Cyclic Interference Alignment (IA) on the \mbox{$3$\,-\,user} \mbox{$X$-\,network} and show that it is infeasible to exactly achieve the upper bound of $\frac{K^2}{2K-1}=\frac{9}{5}$ degrees of freedom for the lower bound of $n=5$ signalling dimensions and $K=3$ user-pairs.
This infeasibility goes beyond the problem of common eigenvectors in invariant subspaces within spatial IA.

In order to gain non-asymptotic feasibility with minimal intervention, we first investigate an alignment strategy that enables IA by feedforwarding a subset of messages with minimal rate.
In a second step, we replace the proposed feedforward strategy by an analogous Cyclic Interference Alignment and Cancellation scheme with a backhaul network on the receiver side and also by a dual Cyclic Interference Neutralization scheme with a backhaul network on the transmitter side.
\end{abstract}

\section{Introduction}
In the seminal work \cite{001}, the basic principle of Interference Alignment (IA) and its consequences for a $K$-user interference channel is introduced.
Besides the thorough discussion of the spatial IA scheme, those authors also briefly described IA in terms of propagation delay for an illustrative example.
In \cite{Z3a}, the given example has been revisited and formulated \wrt a cyclic channel model that operates on cyclically shifted polynomials.
The mathematical framework is inspired by the representation of cyclic codes by polynomial rings in~\cite{592}.
The proposed Cyclic Polynomial Channel Model (CPCM) is closely related to the Linear Deterministic Channel Model (LDCM) introduced by Avestimehr \etal in \cite{005}.
\mbox{Especially} in the case of multiple users exceeding two trans\-mitter-receiver pairs, capacity results in the LDCM are mainly provided for symmetric channel gains, \eg, as in \cite{177,530}. 
A significant property of the LDCM is the linear down-shift of finite dimensional coding vectors.
But this property involves to track a number of side-effects when considering general asymmetric channels.
It is quite challenging to derive optimal communication schemes in closed-form since the number of parameters involved increases exponentially with the total number of users~$K$.
In contrast to such linear shifts, we observed that the use of cyclic shifts is quite beneficial to derive some closed-form solutions with a notable lower complexity for arbitrary channel symmetry. 

An interesting analogy in terms of the feasibility conditions used in \cite{Z3a} is observed in~\cite{569} for a particular \mbox{$3$\,-\,user} inter\-ference channel in an OFDM system with two orthogonal subcarriers.
Moreover, a precoding-based network alignment scheme on a finite-field channel model for the \mbox{$2$\,-\,user} \mbox{$X$\,-\,channel} in \cite{597} is also subject to closely related feasibility conditions.

The combination of Interference Alignment and Cancellation (IAC) is initially introduced in \cite{026} and also applied in~\cite{584}.
Therein, a backhaul network \mbox{(BHN)} provides a limited exchange of messages at the receiver side to support the cancellation of known interference by the aid of other receivers.

Yet another related approach that is important for this work is to inhibit interference by \emph{Interference Neutralization} (IN) \cite{123,350,Z4a}, a communication scheme cancelling interference 'over the air' by aligning complementary versions of the same message within the same signalling space.

\begin{figure}[t]
 \vspace{1mm}
 \centering
 \includegraphics[width=78mm]{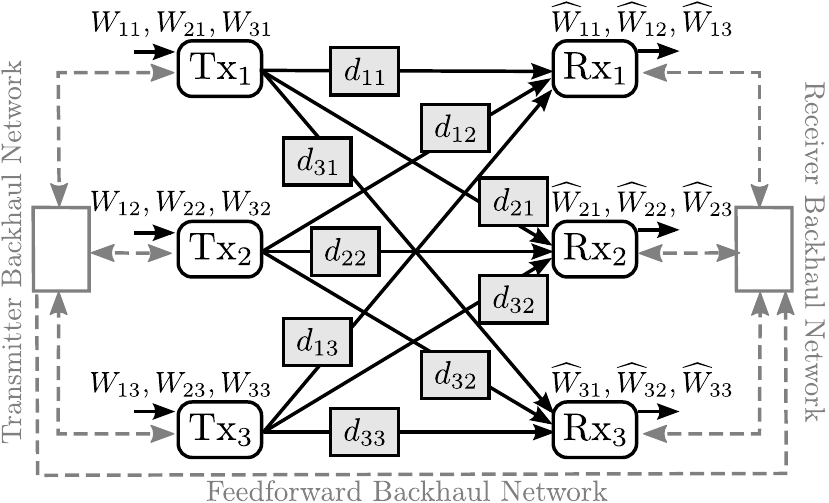}
 \vspace{-1mm}
 \caption{The fully-connected $3$-user $X$-network with $3$ transmitters $\mathrm{Tx}_1, \mathrm{Tx}_2$, $\mathrm{Tx}_3$, $3$ receivers $\mathrm{Rx}_1, \mathrm{Rx}_2,\mathrm{Rx}_3$, $9$ independent messages $W_{ji}$ and $9$ corresponding estimated messages $\widehat{W}_{ji}$.
 The influence of the channel between transmitter $\Tx_i$ and receiver $\Rx_j$ is parameterized by~$d_{ji}$ as indicated by solid black arrows.
 The interference-free backhaul networks for IA, IAC and IN are depicted by dashed grey arrows.}
 \label{fig:channel}
 \vspace{-6mm}
\end{figure}

\textbf{Contributions.}
In this work, we show that perfect Cyclic IA in a $K$\,-\,user $X$\,-\,network for $K \geq 3$ users is overconstraint and hence infeasible.
We observe that this property extends the related problem of common invariant subspaces in spatial~IA~\cite{059b}.
In order to tackle this infeasibility, we first analyze a simple feedforward scheme achieving $\frac{9}{5}$ Degrees of Freedom (DoF) within $5$ dimensions and only one message over the interference-free feedforward~BHN.
Our second step leads us to a related Cyclic IAC scheme on a receiver-sided BHN.
It achieves the $\frac{9}{5}$ DoF with a minimum of only $2$ messages over the BHN.
We observe a duality between IAC on the receiver-sided BHN and IN on the dual transmitter-sided BHNs.
This insight is related to the observations in~\cite{218}.

\textbf{Organization.}
The CPCM of the $K$-user $X$-\,network is presented in Sec.\,\ref{sec:sysmod}.
The infeasibility of perfect Cyclic~IA is shown in Sec.\,\ref{sec:IA-infeasible}.
Cyclic IA with a feedforward BHN, the Cyclic~IAC scheme and the Cyclic~IN scheme are presented in Sec.\,\ref{sec:IAC}.

\section{System Model}
\label{sec:sysmod}

We adapt the notation from \cite{Z3a} and \cite{Z4a} to the given problem.
A $K$-user \emph{$X$-\,channel} describes a wireless channel with $K$~transmitters and $K$~receivers, as depicted in Fig.\,\ref{fig:channel} for $K=3$.
The set of user indices is $\mathcal{K}:=\{1,2, \dots, K\}$.
The $K$-user $X$-channel is physically equivalent to the \mbox{$K$-user} interference channel \cite{001,Z3a}.
But in the $K$-user $X$-network, there are $K^2$ unicast messages with exactly one dedicated message from each transmitter to each receiver.
Such a dedicated message from transmitter $\Tx_i$ to receiver $\Rx_j$, with $i,j \in \mathcal{K}$, is denoted by~$W_{ji}$.

We consider polynomial rings $\mathbb{F}(x)$ modulo $x^n-1$ with the indeterminate~$x$.
The channel access at each $\Tx_i$ and $\Rx_j$ is partitioned into $n \in \mathbb{N}$ equally sized dimensions, each normalized to length~one.
A single {\ts} in the period of $n$ {\ts}s is addressed by one of the \emph{offsets} $x^0, x^1,\dots, x^{n-1}$.
A transmitter $\Tx_i$ can allocate coded messages to each coefficient.
A message is a binary string $W_{ji} \in \mathbb{B}^t=\{0,1\}^t$ with $t \in \mathbb{N}$ symbols.
Altogether, the transmitted signal from $\Tx_i$ is a polynomial with messages $W_{ji} \in \mathbb{B}^t$ for each receiver $\Rx_j$:
\begin{align}
 \label{eqn:transmitsig}
 u_i(x) \equiv \sum\nolimits_{j \in \mathcal{K}} W_{ji} x^{p_{ji}} \modxn.
\end{align}
The parameters $p_{ji} \in \mathbb{N}$ allocate the message $W_{ji}$ to a particular offset.
The influence of the channel is modelled by cyclic shifts of the transmitted polynomials.
A cyclic shift of $k$ offset positions in $u(x)$ is expressed by the multiplication~$x^k u(x)$ modulo $x^n-1$.
E.g., to shift $u(x)=Wx^l$ by~$k$ positions, we compute \mbox{$x^k u(x) \equiv Wx^{l+k}\modxn$} within the period of $n$~\mbox{{\ts}s}.
An arbitrary shift between a pair~($\Rx_j$,$\Tx_i$) is denoted by \mbox{$d_{ji} \in \mathcal{D}=\{x^k| k \in \mathbb{N}\}$}.
The channel matrix is defined by $\boldsymbol{D} = (d_{ji})_{1 \leq j \leq K, 1 \leq i \leq K}$ and it is assumed to be fully known to all users and constant over time.
The received signal $r_j(x)$ at $\Rx_j$ is a superposition of cyclically shifted signals $u_i(x)$:
\begin{align}
 \label{eqn:rec_sig}
 r_j(x) \equiv \sum\nolimits_{i \in \mathcal{K}} d_{ji} u_{i}(x) \modxn.
\end{align}

In the CPCM, the achieved total number of the DoF measures the number of dedicated messages $M$ received interference-free per $n$~{\ts}s \cite{Z3a}:
\begin{align}
 \DoF = \frac{M}{n}. \label{eqn:DoF}
\end{align}
The total number of the DoF is upper bounded for $m_{ji}$ submessages per transmitter-receiver pair $(\Rx_j,\Tx_i)$ by:
\begin{align}
 \label{eqn:u-bound}
 \DoF &\leq \frac{\sum\nolimits_{j=1}^{K_{\mathrm{R}}}\sum\nolimits_{i=1}^{K_{\mathrm{T}}} m_{ji}}
 {\underset{m_{ji}}{\max}\left(\sum\nolimits_{i=1}^{K_{\mathrm{T}}} m_{ji} + \sum\nolimits_{j=1}^{K_{\mathrm{R}}} m_{ji}-m_{ji}\right)}.
\end{align}
as shown in \cite[Sec.\,VI]{Z3a}.
The numerator represents the total number of messages $M$ and the denominator is the lower bound on the number of dimensions $n$.
This upper bound for the CPCM is basically analogous to the bound in \cite[Thm. 1]{059b} but applied to the CPCM-version of a multi-user \mbox{$X$-\,network}.

The bound on the number of DoF is maximal if the $\max$-term in the denominator of \eqref{eqn:u-bound} is the same for all entries~$m_{ji}$.
Only in the symmetric case, i.e, with an equal number of submessages $m_{ji}=L$, $L \in \mathbb{N}$, for all $i,j \in \mathcal{K}$, the upper bound yields its maximum $\frac{K^2}{2K-1}$~DoF, \eg, $\frac{9}{5}$~DoF for $K=3$.

\section{Cyclic IA on the $3$-user $X$-\,Network}
\label{sec:IA-infeasible}
The case for $K=2$ users with the same number of messages $m_{ji}=1$ for all $i,j \in \mathcal{K}$, has already been solved in our previous work \cite[Sec.\,III]{Z3a}.
Here, each transmitter intends to convey \mbox{$K=3$} dedicated messages, one to each receiver.
We have at total number of $M= K^2 = 9$ unicast messages in the system.
Each receiver must decode three dedicated messages, while it must cope with two interfering messages per transmitter.

\subsection{Separability Conditions}
To successfully decode, each dedicated message must be received interference-free.
A message is called interference-free, if three types of \emph{separability conditions} \cite{Z3a} hold for a proper choice of each transmission parameter $p_{ji}$.
Including all $3^2=9$ messages, the \emph{intra-user interference conditions} at $\Tx_i$ for pair-wise distinct $p_{ii},p_{ji},p_{ki}$~are:
\begin{align}
 x^{p_{ji}} & \nequiv x^{p_{ki}} \modxn, \label{eqn:intra-X1} \\
 x^{p_{ii}} & \nequiv x^{p_{ji}} \modxn, \label{eqn:intra-X2} \\
 x^{p_{ii}} & \nequiv x^{p_{ki}} \modxn. \label{eqn:intra-X3}
\end{align}
The \emph{multiple-access interference conditions} at $\Rx_i$ are:
\begin{align}
 d_{ij}x^{p_{ij}} & \nequiv d_{ik}x^{p_{ik}} \modxn,  \label{eqn:mac-X1} \\
 d_{ii}x^{p_{ii}} & \nequiv d_{ij}x^{p_{ij}} \modxn, \label{eqn:mac-X2} \\
 d_{ii}x^{p_{ii}} & \nequiv d_{ik}x^{p_{ik}} \modxn. \label{eqn:mac-X3}
\end{align}
And the \emph{inter-user interference conditions} at $\Rx_i$ are:
\begin{align}
 d_{ii} x^{p_{ii}} & \nequiv d_{ij}x^{p_{kj}} \modxn, \label{eqn:inter-X1} \\
 d_{ii} x^{p_{ii}} & \nequiv d_{ij}x^{p_{jj}} \modxn, \label{eqn:inter-X2} \\
 d_{ii} x^{p_{ii}} & \nequiv d_{ik}x^{p_{jk}} \modxn, \label{eqn:inter-X3} \\
 d_{ii} x^{p_{ii}} & \nequiv d_{ik}x^{p_{kk}} \modxn, \label{eqn:inter-X4} \\
 d_{ij} x^{p_{ij}} & \nequiv d_{ii}x^{p_{ji}} \modxn, \label{eqn:inter-X5} \\
 d_{ij} x^{p_{ij}} & \nequiv d_{ii}x^{p_{ki}} \modxn, \label{eqn:inter-X6} \\
 d_{ij} x^{p_{ij}} & \nequiv d_{ik}x^{p_{jk}} \modxn, \label{eqn:inter-X7} \\
 d_{ij} x^{p_{ij}} & \nequiv d_{ik}x^{p_{kk}} \modxn, \label{eqn:inter-X8} 
\end{align}
for distinct indices $i,j,k \in \mathcal{K}$, respectively.
By a circular relabelling of indices, analogous conditions are expressed for $\Tx_j$, $\Tx_k$, $\Rx_j$ and $\Rx_k$.

\subsection{Infeasibility Problem of Perfect Cyclic~IA}
$\Rx_i$ receives six interfering signals in total: $d_{ii}x^{p_{ji}}$, $d_{ii}x^{p_{ki}}$, $d_{ij}x^{p_{kj}}$, $d_{ij}x^{p_{jj}}$, $d_{ik}x^{p_{jk}}$, $d_{ik}x^{p_{kk}}$.
Two inter\-fering signals from the same transmitter can not be aligned due to the intra-user interference conditions \eqref{eqn:intra-X1} to \eqref{eqn:intra-X3}.
As $3$ dimensions are reserved for dedicated signals and at least $2$ must be reserved for interference, we demand $n \geq 5$.
Perfect Cyclic~IA is optimal and requires exactly~$n=5$.
In the following, we omit that all congruences are reduced modulo $x^5-1$ for brevity.
With perfect Cyclic IA, three interference signals, i.e., one from each transmitter, must be aligned to a single dimension reserved for interference only.
A potential IA scheme at $\Rx_i$ is constructed by choosing one element of each of these three sets as implied by curly~brackets:
 \begin{align}
  \left\{ \hspace{-1mm}
  \begin{array}{l}
   d_{ii}x^{p_{ji}} \\
   d_{ii}x^{p_{ki}}
  \end{array} \hspace{-1.5mm} \right\} \equiv
  \left\{\hspace{-1mm}
  \begin{array}{c}
   d_{ij}x^{p_{jj}} \\
   d_{ij}x^{p_{kj}}
  \end{array}\hspace{-1.5mm} \right\} \equiv
  \left\{\hspace{-1mm}
  \begin{array}{c}
   d_{ik}x^{p_{jk}} \\
   d_{ik}x^{p_{kk}}
  \end{array}\hspace{-1.5mm}\right\}, \label{eqn:IAeqns}
 \end{align}
\eg, by taking the elements in the first row, we obtain: $d_{ii}x^{p_{ji}} \equiv d_{ij}x^{p_{jj}} \equiv  d_{ik}x^{p_{jk}}$ for a fully symmetric perfect alignment scheme similar to \cite[Sec.\,V-C]{059b}.
The IA in one interference dimension directly implicates the complementary alignment of the other interference dimension at the same receiver, \eg, the elements of the second row for the example given above: $d_{ii}x^{p_{ki}} \equiv d_{ij}x^{p_{kj}} \equiv d_{ik}x^{p_{kk}}$.

For notational convenience, we denote submatrices of~$\boldsymbol{D}$~for the computation of minors by:
\begin{align*}
\boldsymbol{D}_{i,k,j,l} =
  \left(
   \begin{array}{cc}
    d_{ij} & d_{il} \\
    d_{kj} & d_{kl}
   \end{array}
  \right).
\end{align*}
Note that the determinant implies the following symmetries:
\begin{align*}
 \det(\boldsymbol{D}_{i,k,j,l}) & \equiv \det(\boldsymbol{D}_{k,i,l,j}) \equiv \\
-\det(\boldsymbol{D}_{k,i,j,l}) & \equiv -\det(\boldsymbol{D}_{i,k,l,j}) \modxn.
\end{align*}

\begin{theorem}
\label{thm:perfIA}
 Perfect Cyclic~IA is infeasible on the \mbox{$3$-user} \mbox{$X$-\,network} with $m_{ij}=1$, for all $i,j \in \mathcal{K}$, and $n=5$.
\end{theorem}
\emph{Proof:}
 Each $\boldsymbol{D}_{i,k,j,l}$ corresponds to a subordinate $2 \times 2$ \mbox{$X$-\,channel} matrix with distinct transmitters $\Tx_j$, $\Tx_l$, and distinct receivers $\Rx_i$, $\Rx_k$.
 Note that for the $2 \times 2$ \mbox{$X$-\,channel}, a non-zero determinant of the channel matrix is necessary to perform IA, as we have already shown in \cite[Thm.\,1\,(a)]{Z3a}.

 Firstly, we assume that $\det(\boldsymbol{D}_{i,j,i,j}) \equiv 0$ holds w.l.o.g. 
 On the one hand, we could align $d_{ii}x^{p_{ji}}\equiv d_{ij}x^{p_{jj}}$ and the above assumption implies $d_{jj}x^{p_{jj}} \equiv d_{ji}x^{p_{ji}}$.
 But this contradicts the mul\-tiple-access interference conditions.
 In analogy, aligning $d_{jj}x^{p_{ij}}\equiv d_{ji}x^{p_{ii}}$ implies that $d_{ii}x^{p_{ii}} \equiv d_{ij}x^{p_{ij}}$, yielding another violation.
 Hence we need $\det(\boldsymbol{D}_{i,j,i,j}) \nequiv 0$ for these schemes.

 Contrariwise, aligning $d_{ii}x^{p_{ki}}\equiv d_{ij}x^{p_{kj}}$ with the initial assumption $\det(\boldsymbol{D}_{i,j,i,j}) \equiv 0$ implies $d_{ji}x^{p_{ki}} \equiv d_{jj}x^{p_{kj}}$.
 This is not a contradiction so far.
 Beyond that, the above assumption is even necessary if both of these alignments are used.

 However, considering the complementary alignment at $\Rx_i$ to \mbox{$d_{ii}x^{p_{ki}} \equiv d_{ij}x^{p_{kj}}$} as given by \eqref{eqn:IAeqns} provides the first alignment scheme $d_{ii}x^{p_{ji}}\equiv d_{ij}x^{p_{jj}}$, demanding \mbox{$\det(\boldsymbol{D}_{i,j,i,j}) \nequiv 0$}.
 Thus, the separability conditions are violated for both $\det(\boldsymbol{D}_{i,j,i,j}) \equiv 0 $ and $\det(\boldsymbol{D}_{i,j,i,j}) \nequiv 0 $ leaving no feasible~$\boldsymbol{D}$.
 This conflict carries over to all minors of $\boldsymbol{D}$ analogously.~$\blacksquare$

Thm.\,\ref{thm:perfIA} entails that perfect Cyclic IA is also infeasible for the $K$-\,user $X$-\,channel with $K \geq 3$ users since there are $K\choose 3$ embedded $3$\,-\,user $X$-\,networks.

Note that this problem does not exclude non-perfect Cyclic IA schemes with $n > 2K-1$ dimensions.
But non-perfect schemes do not achieve the upper bound exactly.

Interestingly, the well-known problem of common eigenvectors in \mbox{invariant} subspaces for perfect spatial IA, as discussed and proven in \cite[\mbox{Sec.\,V-C}]{059b} for $X$-\,networks, is recognizable.
But this problem is only a subordinary part of the infeasibility problem presented in Thm.\,\ref{thm:perfIA} above. 
To briefly elaborate this, we consider a symmetric perfect Cyclic IA scheme which is analogous to the spatial IA scheme in \cite[Eqns.\,(10)-(12)]{059b}:
 \begin{align}
  d_{ii}x^{p_{ji}} &\equiv d_{ij}x^{p_{jj}} \equiv d_{ik}x^{p_{jk}}, \label{eqn:X-IA1} \\
  d_{ii}x^{p_{ki}} &\equiv d_{ij}x^{p_{kj}} \equiv d_{ik}x^{p_{kk}}, \label{eqn:X-IA2}
 \end{align}
 for pair-wise distinct indices~\mbox{$i,j,k \in \mathcal{K}$}.
 Due to symmetry, the alignment of \eqref{eqn:X-IA1} at receiver $\Rx_k$ corresponds to:
 \begin{align}
  d_{kk}x^{p_{jk}} &\equiv d_{kj}x^{p_{jj}} \equiv d_{ki}x^{p_{ji}}, \label{eqn:X-IA3}
 \end{align}
 by a simple relabelling of indices.
 With \eqref{eqn:X-IA1}, \eqref{eqn:X-IA3}, we obtain:
 \begin{align}
	       d_{ii}x^{p_{ji}} &\equiv d_{ij}x^{p_{jj}}, \nonumber \\
	       d_{ki}x^{p_{ji}} &\equiv d_{kj}x^{p_{jj}}, \nonumber \\
  \Rightarrow  d_{ii}d_{ij}^{-1} d_{kj} d_{ki}^{-1} x^{p_{ji}} &\equiv x^{p_{ji}} \label{eqn:cond-perfIA1}.
 \end{align}
An analogous formulation arises in \cite[Eq.\,(15)]{059b} with corresponding diagonal MIMO channel matrices and leads to the problem of common eigenvectors in perfect spatial~IA.
But in the case of Cyclic~IA, the result of \eqref{eqn:cond-perfIA1} only implies the constraint that $\det(\boldsymbol{D}_{i,k,j,i}) \equiv 0$ is needed.

\section{$3$-User $X$-\,Networks with Minimal Backhaul}
\label{sec:IAC}
Since perfect Cyclic~IA is shown to be an overconstrained problem, our approach is to relax a minimal number of conditions by providing a limited number of the messages over a BHN to achieve sufficient feasibility.

\subsection{Cyclic IA with Minimal Feedforward Backhaul Networks}
\label{sec:cyclicIA-FF}
A first and very intuitive approach is to include a wired \emph{feedforward backhaul network} (FF-BHN) between some transmitters and receivers.
A single FF link $\theta_{\mathrm{FF},ji}$ between $\Tx_i$ and $\Rx_j$ with rate $\Theta_{\mathrm{FF},ji}=1$ simply bypasses the channel $d_{ji}$ so that the actual transmission of a forwarded message may be omitted.
The FF-BHN is also shown at the bottom of Fig.\,\ref{fig:channel}.
Note that this approach corresponds to using one cognitive receiver $\Rx_j$ knowing $W_{ji}$.

We propose the following alignment scheme and prove its optimality \wrt the minimal necessary sum-rate $\Theta_{\mathrm{FF}}$:

At~$\Rx_i$, interference from $\Rx_j$ and $\Rx_k$ is perfectly aligned within two dimensions:
\begin{align}
 d_{ii} x^{p_{ji}} &\equiv d_{ij} x^{p_{kj}} \equiv d_{ik} x^{p_{jk}}, \label{eqn:IAC1} \\
 d_{ii} x^{p_{ki}} &\equiv d_{ij} x^{p_{jj}} \equiv d_{ik} x^{p_{kk}}. \label{eqn:IAC2}
\end{align}
The dedicated and interfering signals at $\Rx_j$ are aligned by:
\begin{align}
 d_{ji} x^{p_{ki}} \equiv d_{jj} x^{p_{kj}} &\equiv d_{jk} x^{p_{ik}}, \label{eqn:IAC3}\\
 d_{ji} x^{p_{ii}} &\equiv d_{jk} x^{p_{kk}}, \label{eqn:IAC4}\\
 d_{jj} x^{p_{ij}} &\equiv d_{jk} x^{p_{jk}}, \label{eqn:IAC5:violate1}
\end{align}
and similarly, we use the following Cyclic~IA scheme at $\Rx_k$:
\begin{align}
 d_{ki} x^{p_{ii}} \equiv d_{kj} x^{p_{jj}} &\equiv d_{kk} x^{p_{ik}}, \label{eqn:IAC6}\\
 d_{kj} x^{p_{ij}} &\equiv d_{ki} x^{p_{ji}}, \label{eqn:IAC7} \\
 d_{ki} x^{p_{ki}} &\equiv d_{kk} x^{p_{jk}}. \label{eqn:IAC8:violate2}
\end{align}
A relabelling of indices is not permitted in this asymmetric scheme.
Note that \eqref{eqn:IAC5:violate1} and \eqref{eqn:IAC8:violate2} explicitly violate the separability conditions.
Independent of channel matrix $\boldsymbol{D}$, the dedicated messages $W_{jk}$ and $W_{ki}$ can not be decoded yet.

\begin{theorem}
\label{thm:cyclicIA-FF}
The upper bound of $\frac{9}{5}$ DoF for $n=5$ on the \mbox{$3$-user} $X$\,-\,network is achievable by Cyclic~IA with a FF-BHN and \mbox{$\Theta_{\mathrm{FF}}\geq 1$}.

For the considered scheme in \eqref{eqn:IAC1} to \eqref{eqn:IAC8:violate2}, we assume that:
\begin{itemize}
 \item [(i)]    $d_{ij}d_{ki}d_{jk} \equiv d_{ji}d_{ik}d_{kj}$,
 \item [(ii)]   $d_{ii}d_{jk}d_{kj} \equiv d_{jj}d_{ik}d_{ki} \equiv d_{kk}d_{ij}d_{ji}$,
 \item [(iii)]  $\det(\boldsymbol{D}_{i,k,i,k})\,\nequiv 0$,
 \item [(iv)]   $\det(\boldsymbol{D}_{i,j,i,j})\,\,\nequiv 0$,
 \item [(v)]    $\det(\boldsymbol{D}_{i,j,i,k})\,\nequiv 0$,
 \item [(vi)]   $\det(\boldsymbol{D}_{i,j,j,k})\,\nequiv 0$,
 \item [(vii)]  $\det(\boldsymbol{D}_{k,j,k,i}) \nequiv 0$,
 \item [(viii)] $\det(\boldsymbol{D}_{k,j,k,j}) \nequiv 0$,
 \item [(ix)]   $d_{ii}d_{jj}d_{kk} \nequiv d_{ij}d_{jk}d_{ki} \equiv d_{ji}d_{kj}d_{ik}$,
 \item [(x)]    $d_{kk}d_{ii}d_{jj}d_{kk} \nequiv d_{jk}d_{kj}d_{ik}d_{ki}$, \newline
		$d_{ii}d_{ii}d_{jj}d_{kk}\,\,\nequiv d_{ij}d_{ji}d_{ik}d_{ki}$, \newline
	        $d_{jj}d_{ii}d_{jj}d_{kk}\, \nequiv d_{ij}d_{ji}d_{jk}d_{kj}$, 
\end{itemize}
hold for distinct indices $i,j,k \in \mathcal{K}$.
\end{theorem}
\emph{Proof:} (a) \emph{Necessity of $\Theta_{FF} \geq 1$}: \newline
Since $\Theta_{\mathrm{FF}}=0$ would correspond to using no FF at all, the necessity of $\Theta_{\mathrm{FF}} \geq 1$ follows from Thm.\,\ref{thm:perfIA} evidently.

\begin{figure}[t]
 \vspace{2mm}
 \centering
 \includegraphics[width=67mm]{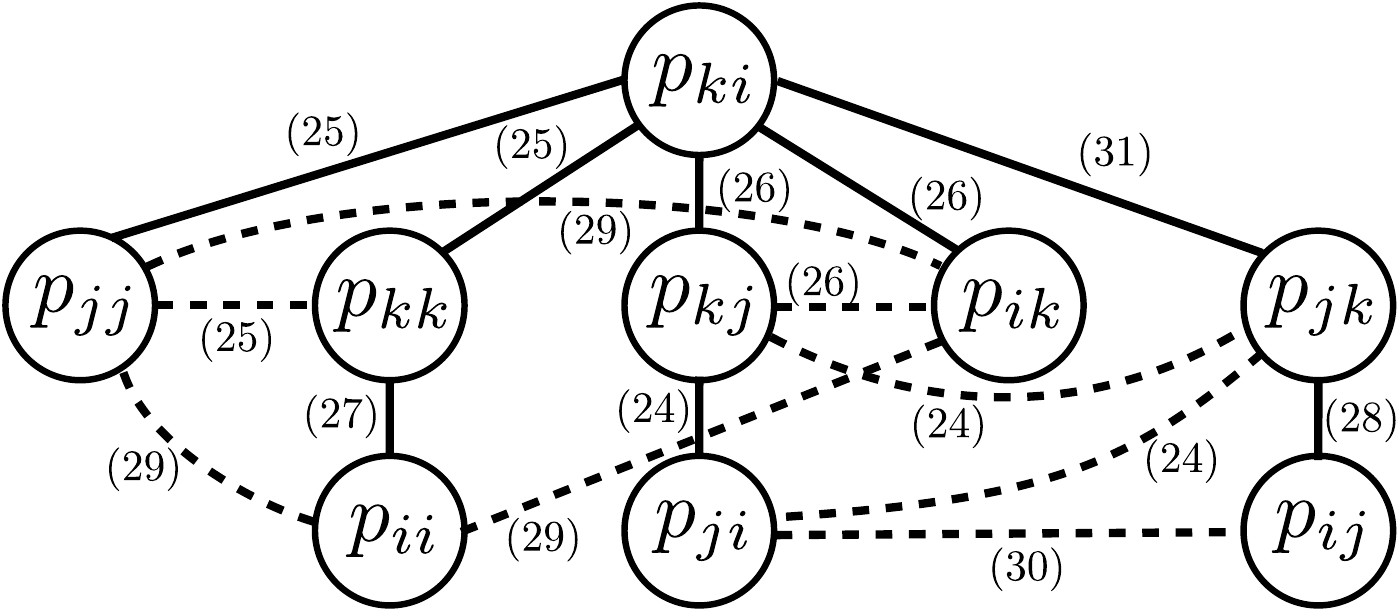}
 \caption{Adjacency graph of the Cyclic~IAC scheme in \eqref{eqn:IAC1} to \eqref{eqn:IAC8:violate2}.
	  Solid lines indicate the assignments for a fixed $p_{ki}$ given in the proof of Thm.\,\ref{thm:cyclicIA-FF}, and dashed lines the remaining conditions that must be checked for feasibility.}
 \label{fig:AdjGraph}
 \vspace{-6mm}
\end{figure}

(b) \emph{Necessity of constraints (i) to (x) for the given scheme}: \newline
Firstly, we consider the constraints (i), (ii) that are implied by the Cyclic~IA scheme given in \eqref{eqn:IAC1} to \eqref{eqn:IAC8:violate2}.
We depict how the parameters are interlinked by the adjacency graph shown in Fig.\,\ref{fig:AdjGraph}.
Other potentially valid alignment schemes will imply a different set of constraints and hence another adjacency graph.

Constraint (i) is obtained by substituting parameters $p_{jj}$, $p_{kk}$, $p_{ii}$ in \eqref{eqn:IAC2}, \eqref{eqn:IAC3}, \eqref{eqn:IAC6}.

One part of constraint (ii) yields from substituting $p_{jk}$, $p_{ji}$, $p_{ij}$ in \eqref{eqn:IAC1}, \eqref{eqn:IAC5:violate1}, \eqref{eqn:IAC4}.
Another part from (ii) yields from substituting $p_{jj}$, $p_{kj}$, $p_{ik}$ in \eqref{eqn:IAC2}, \eqref{eqn:IAC6}, \eqref{eqn:IAC3}, and the last part from substituting $p_{kj}$, $p_{ki}$, $p_{jk}$ in \eqref{eqn:IAC3}, \eqref{eqn:IAC8:violate2},~\eqref{eqn:IAC1}.

Now, we consider the impact of the separability conditions.
Note that the feasibility of \eqref{eqn:IAC5:violate1} and \eqref{eqn:IAC8:violate2} is provided by the FF-BHN, as we will show in part (b), \ie, these particular violations may be excluded.
Constraint (iii) is derived from \eqref{eqn:mac-X3} and \eqref{eqn:IAC6}.
Constraint (iv) is derived by substituting $p_{ki}$, $p_{ji}$, $p_{kj}$ with \eqref{eqn:intra-X1}, \eqref{eqn:IAC1}, \eqref{eqn:IAC3}.

\iftoggle{EXTENDED}
{The remaining constraints (v) to (x) are proven analogously \wrt all separability conditions at each receiver.
The complete proof of those constraints is provided in Appx.\,\ref{appx:proof-constraints}.}    
{The complete proof of the remaining constraints (v) to (x) \wrt all separability conditions at each receiver is provided in the extended version~\cite{Z8a} of this paper.}		 
The three constraints of (x) are equivalent due to~(ii).

(b) \emph{Sufficiency of Cyclic IA with $\Theta_{\mathrm{FF}}=1$}: \newline
It suffices to bypass the transmission of the dedicated message $W_{jk}$ over $d_{ji}$ through a FF link $\theta_{\mathrm{FF},jk}$ with rate \mbox{$\Theta_{FF}=1$}.
Then, $\Tx_k$~may omit the transmission of~$W_{jk}$ over~$\boldsymbol{D}$ and can still decode $W_{jk}$ from the FF-BHN.
As $W_{jk}$ is not transmitted over $\boldsymbol{D}$ at all, $\Rx_k$ can also decode $W_{ki}$ interference-free.

To show that the proposed IA scheme with FF is feasible now, all nine transmission parameters must be resolved.
As indicated by the adjacency graph in Fig.~\ref{fig:AdjGraph}, we fix the top-most parameter $p_{ki}$, w.l.o.g.
With \eqref{eqn:IAC2}, we obtain $p_{jj},p_{kk}$, \eqref{eqn:IAC3} provides $p_{kj},p_{ik}$, and \eqref{eqn:IAC8:violate2} yields $p_{jk}$.
With \eqref{eqn:IAC4}, $p_{ii}$ yields from $p_{kk}$.
With \eqref{eqn:IAC1}, $p_{ji}$ yields from $p_{kj}$.
And with \eqref{eqn:IAC5:violate1}, $p_{ij}$ yields from $p_{jk}$.

A valid matrix, normalized \wrt the main diagonal is, \eg:
\begin{align*}
  \boldsymbol{D} &= \left(
  \begin{array}{ccc}
   1   & x^4 & x^2 \\
   x^4 & 1   & x^2 \\
   x   & x   & 1
  \end{array} \right),
 \end{align*}
as all constraints (i) $\equiv x^2$, (ii) $\equiv x^3$, (iii) $\equiv 1-x^3$, \mbox{(iv) $\equiv 1-x^3$}, \mbox{(v) $\equiv x^2 -1$}, (vi) $\equiv x^1-x^2$, (vii) $\equiv x^4-x^3$, (viii) $\equiv 1-x^3$, (ix) $1 \nequiv x^2$, (x) $1 \nequiv x^1$ for $i=1$, $j=2$, $k=3$, are fulfilled.
A valid set of transmission parameters \iftoggle{EXTENDED}{\hspace{-1mm}}{\cite[Appx. B]{Z8a}} satisfying all conditions with a fixed $p_{31}=4$~yields the following vector~$\boldsymbol{p}$:
\begin{align*}
 \boldsymbol{p} = &\ (  p_{11},    p_{21},    p_{31},   p_{12},    p_{22},    p_{32},   p_{13},    p_{23},    p_{33}) \\ =&\
    (\ \ \, 0, \ \ \,  2, \ \ \,  4,\ \ \,  2, \ \ \,  0, \ \ \,  3,\ \ \,  1, \ \ \,  0, \ \ \,  2).
\end{align*}
Altogether, $\frac{9}{5}$ DoF are achieved by Cyclic IA \iftoggle{EXTENDED}{(cf. Appx.\,\ref{appx:exmpl})}{\hspace{-1mm}}.~$\blacksquare$

Note that a delayed transmission on the FF-BHN only delays the decoding time but does not affect the feasibility.

We would like to emphasize, that the given constraints on $\boldsymbol{D}$ are profoundly interdependent with the proposed IA scheme.
Nonetheless the analysis of comparable valid Cyclic~IA schemes can be performed analogously to~Thm.\,\ref{thm:cyclicIA-FF}.

\subsection{Cyclic IAC over Minimal Receiver Backhaul Networks}
\label{sec:min-iac-R-BHN}
Now, instead of a FF-BHN, we consider a \emph{receiver backhaul network} \mbox{(R-BHN)}.
The R-BHN only permits that receivers may exchange messages to resolve leaking interference in order to satisfy all separability conditions.
The R-BHN is depicted in Fig.\,\ref{fig:channel} on the right hand side.
A single link with rate $\Theta_{\mathrm{R},{ij}}$ in the R-BHN from $\Rx_i$ to $\Rx_j$ is denoted by~$\theta_{\mathrm{R},ji}$.

Similar to the previous section, our aim is to characterize the minimal sum-rate $\Theta_{\mathrm{R}}$ on the \mbox{R-BHN}, that is necessary to achieve the upper bound of $\frac{9}{5}$~DoF on the $3$\,-\,user $X$-\,network.

\begin{lemma}
 \label{lem:IAC-R-BHN}
 The upper bound of $\frac{9}{5}$ DoF for $n=5$ on the \mbox{$3$\,-\,user} \mbox{$X$-\,network} is achievable by Cyclic IAC with $\Theta_{\mathrm{R}} \geq 2$ for the conditions given in Thm.\,\ref{thm:cyclicIA-FF}.
\end{lemma}

\emph{Proof:}
(a) \emph{Necessity of $\Theta_{\mathrm{R}} \geq 2$:} $\newline$
As Cyclic IA without cancellation is precluded by Thm.~\ref{thm:perfIA} for $n=5$, it follows that $\Theta_{\mathrm{R}}>0$.
In contrast to Thm.~\ref{thm:cyclicIA-FF}, no message can be neglected and bypassed so that all must be sent over the channel $\boldsymbol{D}$.
Thus the case $\Theta_{\mathrm{R}}=1$ demands that the interference at two receivers, say, $\Tx_i$ and $\Tx_j$, must be perfectly aligned and only one interfering signal may leak at the remaining receiver $\Tx_k$.
However, simultaneous perfect Cyclic IA at two receivers is already precluded by Thm.~\ref{thm:perfIA}.

(b) \emph{Sufficiency of Cyclic IAC for $\Theta_{\mathrm{R}}=2$:} $\newline$
We consider the same Cyclic IA scheme as provided in \eqref{eqn:IAC1} to \eqref{eqn:IAC8:violate2} subject to constraints (i) to (x) of Thm.\,\ref{lem:IAC-R-BHN}.
But now, the leaking interference in \eqref{eqn:IAC5:violate1} and \eqref{eqn:IAC8:violate2} is resolved by \mbox{$\Theta_{\mathrm{R}}=2$} messages over the R-BHN.
In particular  $W_{ij}$ is conveyed over $\theta_{\mathrm{R},ji}$, so that $W_{ij}$ can be cancelled from the aligned \mbox{$W_{ij} + W_{jk}$} to decode the dedicated message $W_{jk}$ at $\Rx_j$.
In a subsequent step, $W_{jk}$ is conveyed over $\theta_{\mathrm{R},kj}$, so that $W_{jk}$ is cancelled from the aligned \mbox{$W_{jk} + W_{ki}$} to decode the dedicated message $W_{ki}$ at~$\Rx_k$.
$\blacksquare$

A delayed R-BHN transmission does not affect feasibility as long as the backhaul messages $W_{ij}$ and $W_{jk}$ for cancellation adhere to the proposed sequence.

\subsection{Cyclic IAC over Minimal Transmitter Backhaul Networks}
\label{sec:min-iac-T-BHN}
We now consider the reversed case: Transmitters are connected via a \emph{transmitter backhaul network} (T-BHN) instead.
A backhaul link from $\Tx_i$ to $\Tx_j$ is described by $\theta_{\mathrm{T},ji}$ correspondingly.
The sum-rate over the T-BHN is denoted by $\Theta_{\mathrm{T}}$.
The T-BHN is depicted in Fig.\,\ref{fig:channel} on the right hand side.

\begin{lemma}
 \label{lem:IN-T-BHN}
 The upper bound of $\frac{9}{5}$ DoF for $n = 5$ on the \mbox{$3$\,-\,user} \mbox{$X$-\,network} is achievable by Cyclic IN with $\Theta_{\mathrm{T}} \geq 2$ for the conditions given in Thm.\,\ref{thm:cyclicIA-FF}.
\end{lemma}
\emph{Proof:}
 This is a dual scheme to Lem.\,\ref{lem:IAC-R-BHN} for the \mbox{R-BHN} considered above so that the necessity of $\Theta_{\mathrm{T}} \geq 2$ is analogous.
 Again, we use the alignment scheme of \eqref{eqn:IAC1} to \eqref{eqn:IAC8:violate2} subject to constraints (i) to~(x) of Thm.\,\ref{thm:cyclicIA-FF}.
 But in contrast to Lem.\,\ref{lem:IAC-R-BHN}, $W_{ij}$ is firstly conveyed over $\theta_{\mathrm{T},kj}$ with $\Theta_{\mathrm{T},kj}=1$ and then the combined message $W_{ij}-W_{jk}$ is conveyed over $\theta_{\mathrm{T},ik}$ with $\Theta_{\mathrm{T},ik}=1$. 
 $\Tx_k$ transmits the superposition $W_{jk}-W_{ij}$ instead of $W_{jk}$ only, and $\Tx_i$ transmits the superposition $W_{ki}+W_{ij}-W_{jk}$ instead of $W_{ki}$ only.
 This change does not only maintain the decodability of the dedicated signals received at $\Rx_i$ or $\Rx_j$, but rather neutralizes the previously leaking interference observed at $\Rx_j$ and $\Rx_k$.~$\blacksquare$

In contrast to Cyclic IAC over the R-BHN, the exchange of signals over the T-BHN must be performed at any time before the actual transmission.
This relationship also endorses a related IAC-IN duality property reported in~\cite{218}.

\subsection{Combined Cyclic IAC and IN over Minimal T/R-BHNs}
\label{sec:min-iac-TR-BHN}
If transmitters and receivers are each connected to a disjoint \mbox{T-BHN} and \mbox{R-BHN}, the IAC and IN schemes of Sec.~\ref{sec:min-iac-R-BHN} and~\ref{sec:min-iac-T-BHN} can be combined.
The sum-rate over both BHNs is denoted by $\Theta_{\mathrm{TR}}=\Theta_{\mathrm{R}}+\Theta_{\mathrm{T}}$.
\begin{corollary}
 \label{cor:BHN-3}
 The upper bound of $\frac{9}{5}$ DoF for $n=5$~on the \mbox{$3$-user} $X$-network is achievable by combined Cyclic~IAC and Cyclic~IN with~\mbox{$\Theta_{\mathrm{TR}} \geq 2$} for the conditions given in Thm.\,\ref{thm:cyclicIA-FF}.
\end{corollary}
\emph{Proof}:
 Using \eqref{eqn:IAC1} to \eqref{eqn:IAC8:violate2}, $W_{ij}$ is provided over $\theta_{\mathrm{T},kj}$ and $W_{jk}\hspace{-0.5mm}-\hspace{-0.5mm}W_{ij}$ replaces $W_{ij}$ at $\Tx_k$ so that $W_{ij}$ neutralized at $\Rx_j$.
 $\Rx_k$ receives $W_{ki}\hspace{-0.5mm}+\hspace{-0.5mm}W_{jk}\hspace{-0.5mm}-\hspace{-0.5mm}W_{ij}$. 
 Then, $W_{jk}\hspace{-0.5mm}+\hspace{-0.5mm}W_{ji}$ is provided over $\theta_{\mathrm{R},kj}$.
 $\Rx_k$ decodes $(W_{ki}\hspace{-0.5mm}+\hspace{-0.5mm}W_{jk}\hspace{-0.5mm}-\hspace{-0.5mm}W_{ij})\hspace{-0.5mm}-\hspace{-0.5mm}(W_{jk}\hspace{-0.5mm}+\hspace{-0.5mm}W_{ji})\hspace{-0.5mm}+\hspace{-0.5mm}(W_{ji}\hspace{-0.5mm}+\hspace{-0.5mm}W_{ij})=W_{ki}$ using its interfered signal and~\eqref{eqn:IAC7}.~$\blacksquare$

\vspace{-1mm}
\begin{appendix}
\begin{figure*}[!t]
  \centering
  \begin{tabular}{|l||l|l|l|l|l|}
  \hline
    &$x^0$ & $x^1$& $x^2$& $x^3$& $x^4$ \\ \hline\hline
    $v_1(x)$ &$W_{11}$&$0$&$W_{21}$&$0$&$W_{31}$ \\ \hline
    $v_2(x)$ &$W_{22}$&$0$&$W_{12}$&$W_{32}$&$0$ \\ \hline
    $v_3(x)$ &$W_{23}$&$W_{13}$ &$W_{33}$&$0$& $0$ \\ \hline \hline
    $r_1(x)$ &$W_{11}$&$W_{12}$&$W_{21}\hspace{-0.5mm}+\hspace{-0.5mm}W_{32}\hspace{-0.5mm}+\hspace{-0.5mm}W_{23}$&$W_{13}$&$W_{31}\hspace{-0.5mm}+\hspace{-0.5mm}W_{22}\hspace{-0.5mm}+\hspace{-0.5mm}W_{33}$ \\ \hline
    $r_2(x)$ &$W_{22}$&$ W_{21}$&$W_{23}\hspace{-0.5mm}+\hspace{-0.5mm}W_{12}$&$W_{31}\hspace{-0.5mm}+\hspace{-0.5mm}W_{32}\hspace{-0.5mm}+\hspace{-0.5mm}W_{13}$&$W_{11}\hspace{-0.5mm}+\hspace{-0.5mm}W_{33}$ \\ \hline
    $r_3(x)$ &$W_{31}\hspace{-0.5mm}+\hspace{-0.5mm}W_{23}\phantom{\hspace{-0.5mm}+\hspace{-0.5mm}W_{00}}$&$W_{11}\hspace{-0.5mm}+\hspace{-0.5mm}W_{22}\hspace{-0.5mm}+\hspace{-0.5mm}W_{13}$&$W_{33}$&$W_{21}\hspace{-0.5mm}+\hspace{-0.5mm}W_{12}$&$W_{32}$ \\ \hline
  \end{tabular}
  \caption{Transmitted signals $v_i(x)$ and received signals $r_i(x)$, $i \in \mathcal{K}$, using the channel matrix~$\boldsymbol{D}$ given in \eqref{eqn:D-theta-2}. Cyclic right-shifts are used here.}
  \label{tab:FF}
\end{figure*}
\subsection{Examples for the Given Cyclic IA, IAC and IN Schemes}
\label{appx:exmpl}
\vspace{-2mm}
For all examples we use the valid channel matrix:
\begin{align}
 \label{eqn:D-theta-2}
  \boldsymbol{D} &= \left(
  \begin{array}{ccc}
   1   & x^4 & x^2 \\
   x^4 & 1   & x^2 \\
   x   & x   & 1
  \end{array} \right),
 \end{align}
and we fix the indices $i=1, j=2, k=3$ and the parameter vector $\boldsymbol{p}=(0,2,4,2,0,3,1,0,2)$, as in part (b) of the proof for Thm.\,\ref{thm:cyclicIA-FF}.
The resulting transmitted and received signals for this example is depicted in the table of Fig.~\ref{tab:FF}.
It is easy to see that all dedicated messages are decodable at $\Rx_1, \Rx_2$ and $\Rx_3$, except the messages $W_{23}$ at $\Rx_2$ and $W_{31}$ at $\Rx_3$.

\subsubsection{Cyclic IAC with a FF-BHN}
\label{appx:exmp:IA-FF}
According to Thm.~\ref{thm:cyclicIA-FF}, $\Tx_3$ provides $W_{23}$ to $Rx_2$ via the FF-BHN, so that $W_{23}$ is decodable at $\Rx_2$ now.
As the message $W_{23}$ is not transmitted over the channel $\boldsymbol{D}$ at all $W_{31}$ is finally decodable at $\Rx_3$.

\subsubsection{Cyclic IAC with a R-BHN}
\label{appx:exmp:IAC-R-BHN}
According to Lem.~\ref{lem:IAC-R-BHN}, $\Rx_1$ provides $W_{12}$ to $\Rx_2$ via the R-BHN, so that $W_{32}$ is decodable at $\Rx_2$ now.
After decoding $W_{32}$ at $\Rx_2$, $\Rx_2$ provides $W_{32}$ to $\Rx_3$ via the R-BHN, so that $W_{31}$ is decodable at $\Rx_3$.

\subsubsection{Cyclic IN with a T-BHN}
\label{appx:exmp:IN-T-BHN}
According to Lem.\,\ref{lem:IN-T-BHN}, $\Tx_2$ provides $W_{12}$ to $\Tx_3$ via the T-BHN, and then $\Tx_3$ provides the combined message $W_{12}-W_{23}$ to $\Tx_1$ before the actual transmission of these messages over channel $\boldsymbol{D}$.
$\Tx_3$ transmits $W_{23}-W_{12}$ instead of $W_{23}$ and $\Tx_1$ transmits $W_{31}+W_{23}-W_{12}$ instead of $W_{31}$.
As a result, all interfering messages neutralize each other and all dedicated messages are decodable.

\subsubsection{Cyclic IAC and IN with T/R-BHNs}
\label{appx:exmp:IAC-IN-TR-BHN}
According to Coroll.~\ref{cor:BHN-3}, $\Rx_2$ provides $W_{23}$ to $\Rx_3$ over the T-BHN before transmission.
$\Tx_3$ transmits $W_{23}-W_{12}$ instead of $W_{23}$.
On the receiver side, $\Rx_2$ can decode $W_{23}$ as the interference is neutralized.
Then $\Rx_2$ provides the combined message $W_{23}+W_{21}$ to $\Rx_3$.
As $\Rx_3$ receives $W_{31}+W_{23}-W_{12}$, it adds $W_{21}+W_{12}$ and substracts $W_{23}+W_{21}$ to finally decode~$W_{31}$.

\subsection{Proof of the Constraints in Thm.\,\ref{thm:cyclicIA-FF}}
\label{appx:proof-constraints}
In the following, we prove all that all separability conditions of Thm.\,\ref{thm:cyclicIA-FF} hold.
The two particular exceptions given by \eqref{eqn:IAC5:violate1} and \eqref{eqn:IAC8:violate2} are neglected.
Congruences are taken modulo $x^5-1$.
\vspace{-7mm}
\begin{align*}
	 \eqref{eqn:intra-X1}: x^{p_{ki}} &\nequiv x^{p_{ji}} \\
 \smash[t]{\stackrel{\eqref{eqn:IAC1}}{\Rightarrow}} x^{p_{ki}} &\nequiv d_{ij}d_{ii}^{-1}x^{p_{kj}} \\
 \smash[t]{\stackrel{\eqref{eqn:IAC3}}{\Rightarrow}} x^{p_{ki}} &\nequiv d_{ij}d_{ii}^{-1}d_{ji}d_{jj}^{-1}x^{p_{ki}} \\
 \Rightarrow 0 &\nequiv \det(\boldsymbol{D}_{i,j,i,j}) \Rightarrow \text{(iv)} \\
\vspace{2mm}
	 \eqref{eqn:intra-X2}: x^{p_{ji}} &\nequiv x^{p_{ii}} \\
 \smash[t]{\stackrel{\eqref{eqn:IAC6}}{\Rightarrow}} x^{p_{ji}} &\nequiv d_{kk}d_{ki}^{-1}x^{p_{ik}} \\
 \smash[t]{\stackrel{\eqref{eqn:IAC3}}{\Rightarrow}} x^{p_{ji}} &\nequiv d_{kk}d_{ki}^{-1}d_{jj}d_{jk}^{-1}x^{p_{kj}} \\
 \Rightarrow \smash[t]{\stackrel{\eqref{eqn:IAC1}}{\Rightarrow}} x^{p_{ji}} &\nequiv d_{kk}d_{ki}^{-1}d_{jj}d_{jk}^{-1}d_{ii}d_{ij}^{-1}x^{p_{ji}} \\
 \Rightarrow d_{ii}d_{jj}d_{kk} &\nequiv d_{ij}d_{jk}d_{ki} \Rightarrow \text{(ix)} \\
\vspace{2mm}
	 \eqref{eqn:intra-X3}: x^{p_{ki}} &\nequiv x^{p_{ii}} \\
 \smash[t]{\stackrel{\eqref{eqn:IAC1}}{\Rightarrow}} x^{p_{ki}} &\nequiv d_{jk}d_{ji}^{-1}x^{p_{kk}} \\
 \smash[t]{\stackrel{\eqref{eqn:IAC3}}{\Rightarrow}} x^{p_{ki}} &\nequiv d_{jk}d_{ji}^{-1}d_{ii}d_{ik}^{-1}x^{p_{ki}} \\
 \Rightarrow 0 &\nequiv \det(\boldsymbol{D}_{i,j,i,k}) \Rightarrow \text{(v)} \\
\vspace{2mm}
     \eqref{eqn:mac-X1}: d_{ij}x^{p_{ij}} &\nequiv d_{ik}x^{p_{ik}} \\
 \smash[t]{\stackrel{\eqref{eqn:IAC3}}{\Rightarrow}} x^{p_{ij}} &\nequiv d_{ik}d_{ij}^{-1}d_{jj}d_{jk}^{-1}x^{p_{kj}} \\
 \smash[t]{\stackrel{\eqref{eqn:IAC1}}{\Rightarrow}} x^{p_{ij}} &\nequiv d_{ik}d_{ij}^{-1}d_{jj}d_{jk}^{-1}d_{ii}d_{ij}^{-1}x^{p_{ji}} \\
 \smash[t]{\stackrel{\eqref{eqn:IAC7}}{\Rightarrow}} x^{p_{ij}} &\nequiv d_{ik}d_{ij}^{-1}d_{jj}d_{jk}^{-1}d_{ii}d_{ij}^{-1}d_{kj}d_{ki}^{-1}x^{p_{ij}} \\
 \smash[t]{\stackrel{(i)}{\Rightarrow}} 1 &\nequiv d_{ji}d_{kj}d_{ik}d_{ij} d_{ii}^{-1}d_{jj}^{-1}d_{ik}^{-1}d_{kj}^{-1} \\
 \Rightarrow 0 &\nequiv \det(\boldsymbol{D}_{i,j,i,j}) \Rightarrow \text{(iv)} \\
\vspace{2mm}
     \eqref{eqn:mac-X2}: d_{ii}x^{p_{ii}} &\nequiv d_{ij}x^{p_{ij}} \\
 \smash[t]{\stackrel{\eqref{eqn:IAC5:violate1}}{\Rightarrow}} x^{p_{ii}} &\nequiv d_{ij}d_{ii}^{-1}d_{jk}d_{jj}^{-1}x^{p_{jk}} \\
 \smash[t]{\stackrel{\eqref{eqn:IAC8:violate2}}{\Rightarrow}} x^{p_{ii}} &\nequiv d_{ij}d_{ii}^{-1}d_{jk}d_{jj}^{-1}d_{ki}d_{kk}^{-1}x^{p_{ki}} \\
\smash[t]{\stackrel{\eqref{eqn:IAC2}}{\Rightarrow}} x^{p_{ii}} &\nequiv d_{ij}d_{ii}^{-1}d_{jk}d_{jj}^{-1}d_{ki}d_{kk}^{-1}d_{ik}d_{ii}^{-1}x^{p_{kk}} \\
\smash[t]{\stackrel{\eqref{eqn:IAC4}}{\Rightarrow}} x^{p_{ii}} &\nequiv d_{ij}d_{ii}^{-1}d_{jk}d_{jj}^{-1}d_{ki}d_{kk}^{-1}d_{ik}d_{ii}^{-1}d_{ji}d_{jk}^{-1}x^{p_{ii}} \\
\Rightarrow & d_{ii}d_{jj}d_{kk}d_{ii} \nequiv d_{ij}d_{ji}d_{ik}d_{ki} \Rightarrow \text{(x)} \\
\vspace{2mm}
     \eqref{eqn:mac-X3}: d_{ii}x^{p_{ii}} &\nequiv d_{ik}x^{p_{ik}} \\
 \smash[t]{\stackrel{\eqref{eqn:IAC6}}{\Rightarrow}} x^{p_{ii}} &\nequiv d_{ik}d_{ii}^{-1}d_{ki}d_{kk}^{-1}x^{p_{ii}} \\
 \Rightarrow 0 & \nequiv \det(\boldsymbol{D}_{i,j,i,k}) \Rightarrow \text{(v)} \\
  \eqref{eqn:inter-X1}: d_{ii} x^{p_{ii}} &\nequiv d_{ij}x^{p_{kj}} \modxn, \\
 \smash[t]{\stackrel{\eqref{eqn:IAC3}}{\Rightarrow}} x^{p_{ii}} &\nequiv d_{ij}d_{ii}^{-1}d_{jk}d_{jj}^{-1}x^{p_{ik}} \\
 \smash[t]{\stackrel{\eqref{eqn:IAC6}}{\Rightarrow}} x^{p_{ii}} &\nequiv d_{ij}d_{ii}^{-1}d_{jk}d_{jj}^{-1}d_{ki}d_{kk}^{-1}x^{p_{ii}} \\
	   \Rightarrow d_{ii}d_{jj}d_{kk} &\nequiv d_{ij}d_{jk}d_{ki} \Rightarrow \text{(ix)} \\
  \eqref{eqn:inter-X2}: d_{ii} x^{p_{ii}} &\nequiv d_{ij}x^{p_{jj}} \\
 \smash[t]{\stackrel{\eqref{eqn:IAC6}}{\Rightarrow}} x^{p_{ii}} &\nequiv d_{ij}d_{ii}^{-1}d_{ki}d_{kj}^{-1}x^{p_{ii}} \\
 \Rightarrow 0 & \nequiv \det(\boldsymbol{D}_{i,j,i,k}) \Rightarrow \text{(v)}
\end{align*}
\begin{align*}
 \eqref{eqn:inter-X3}: d_{ii} x^{p_{ii}} & \nequiv d_{ik}x^{p_{jk}} \\
 \smash[t]{\stackrel{\eqref{eqn:IAC4}}{\Rightarrow}} x^{p_{jk}} &\nequiv d_{ii}d_{ik}^{-1}d_{jk}d_{ji}^{-1}x^{p_{kk}} \\
 \smash[t]{\stackrel{\eqref{eqn:IAC2}}{\Rightarrow}} x^{p_{jk}} &\nequiv d_{ii}d_{ik}^{-1}d_{jk}d_{ji}^{-1}d_{ii}d_{ik}^{-1}x^{p_{kk}} \\
 \smash[t]{\stackrel{\eqref{eqn:IAC8:violate2}}{\Rightarrow}} x^{p_{jk}} &\nequiv  d_{ii}d_{ik}^{-1}d_{jk}d_{ji}^{-1}d_{ii}d_{ik}^{-1}d_{kk}d_{ki}^{-1}x^{p_{jk}} \\
\smash[t]{\stackrel{(ii)}{\Rightarrow}} d_{ii}d_{jj}d_{kk} &\nequiv d_{ik}d_{ji}d_{kj} \Rightarrow \text{(ix)} \\
  \eqref{eqn:inter-X4}: d_{ii} x^{p_{ii}} &\nequiv d_{ik}x^{p_{kk}} \\
 \smash[t]{\stackrel{\eqref{eqn:IAC4}}{\Rightarrow}} x^{p_{ii}} &\nequiv d_{ik}d_{ii}^{-1}d_{ji}d_{jk}^{-1}x^{p_{ii}} \\
 \Rightarrow 0 & \nequiv \det(\boldsymbol{D}_{i,j,i,k}) \Rightarrow \text{(v)} \\
  \eqref{eqn:inter-X5}: d_{ij} x^{p_{ij}} &\nequiv d_{ii}x^{p_{ji}} \\
 \smash[t]{\stackrel{\eqref{eqn:IAC7}}{\Rightarrow}} x^{p_{ij}} &\nequiv d_{ii}d_{ij}^{-1}d_{kj}d_{ki}^{-1}x^{p_{ij}} \\
 \Rightarrow 0 &\nequiv \det(\boldsymbol{D}_{i,j,i,k}) \Rightarrow \text{(v)} \\
	   \eqref{eqn:inter-X6}: d_{ij} x^{p_{ij}} &\nequiv d_{ii}x^{p_{ki}} \\
 \smash[t]{\stackrel{\eqref{eqn:IAC8:violate2}}{\Rightarrow}} x^{p_{ij}} &\nequiv d_{ii}d_{ij}^{-1}d_{kk}d_{ki}^{-1}x^{p_{jk}} \\
 \smash[t]{\stackrel{\eqref{eqn:IAC5:violate1}}{\Rightarrow}} x^{p_{ii}} &\nequiv d_{ii}d_{ij}^{-1}d_{kk}d_{ki}^{-1}d_{jj}d_{jk}^{-1}x^{p_{ij}} \\
	 \Rightarrow d_{ii}d_{jj}d_{kk} &\nequiv d_{ij}d_{jk}d_{ki} \Rightarrow \text{(ix)} \\
	   \eqref{eqn:inter-X7}: d_{ij} x^{p_{ij}} &\nequiv d_{ik}x^{p_{jk}} \\
 \smash[t]{\stackrel{\eqref{eqn:IAC5:violate1}}{\Rightarrow}} x^{p_{ij}} &\nequiv d_{ik}d_{ij}^{-1}d_{jj}d_{jk}^{-1}x^{p_{ij}} \\
 \Rightarrow 0 &\nequiv \det(\boldsymbol{D}_{i,j,j,k}) \Rightarrow \text{(vi)} \\
 \eqref{eqn:inter-X8}: d_{ij} x^{p_{ij}} &\nequiv d_{ik}x^{p_{kk}} \\
 \smash[t]{\stackrel{\eqref{eqn:IAC5:violate1}}{\Rightarrow}} x^{p_{ij}} &\nequiv d_{ij}d_{ik}^{-1}d_{jk}d_{jj}^{-1}x^{p_{jk}} \\
 \smash[t]{\stackrel{\eqref{eqn:IAC8:violate2}}{\Rightarrow}} x^{p_{ij}} &\nequiv d_{ij}d_{ik}^{-1}d_{jk}d_{jj}^{-1}d_{ki}d_{kk}^{-1}x^{p_{ki}} \\
 \smash[t]{\stackrel{\eqref{eqn:IAC2}}{\Rightarrow}} x^{p_{ij}} &\nequiv  d_{ij}d_{ik}^{-1}d_{jk}d_{jj}^{-1}d_{ki}d_{kk}^{-1}d_{ik}d_{ii}^{-1}x^{p_{kk}} \\
\Rightarrow d_{ii}d_{jj}d_{kk} &\nequiv d_{ij}d_{jk}d_{ki} \Rightarrow \text{(ix)}
\end{align*}
Analogously, we consider the cyclically relabelled versions $i \rightarrow j \rightarrow k \rightarrow i$ of the separability conditions \eqref{eqn:intra-X1}$^\dagger$ to \eqref{eqn:inter-X8}$^\dagger$, with superscript symbol $^\dagger$ denoting the relabelled versions.
\begin{align*}
 \eqref{eqn:intra-X1}^\dagger: x^{p_{kj}} & \nequiv x^{p_{ij}} \\
 \smash[t]{\stackrel{\eqref{eqn:IAC1}}{\Rightarrow}} x^{p_{ij}} &\nequiv d_{ii}d_{ij}^{-1}x^{p_{ji}} \\
 \smash[t]{\stackrel{\eqref{eqn:IAC1}}{\Rightarrow}} x^{p_{ij}} &\nequiv d_{ii}d_{ij}^{-1}d_{kj}d_{ki}^{-1}x^{p_{ij}} \\
 \Rightarrow 0 &\nequiv \det(\boldsymbol{D}_{i,j,i,k}) \Rightarrow \text{(v)} \\
        \eqref{eqn:intra-X2}^\dagger: x^{p_{jj}} &\nequiv x^{p_{kj}} \\
 \smash[t]{\stackrel{\eqref{eqn:IAC2}}{\Rightarrow}} x^{p_{kj}} &\nequiv d_{ii}d_{ij}^{-1}x^{p_{ki}} \\
 \smash[t]{\stackrel{\eqref{eqn:IAC3}}{\Rightarrow}} x^{p_{kj}} &\nequiv d_{ii}d_{ij}^{-1}d_{jj}d_{ji}^{-1}x^{p_{kj}} \\
 \Rightarrow 0 &\nequiv \det(\boldsymbol{D}_{i,j,i,j}) \Rightarrow \text{(iv)}
\end{align*}
\begin{align*}
 \eqref{eqn:intra-X3}^\dagger: x^{p_{jj}} &\nequiv x^{p_{ij}} \\
 \smash[t]{\stackrel{\eqref{eqn:IAC5:violate1}}{\Rightarrow}} x^{p_{jj}} &\nequiv d_{jk}d_{jj}^{-1}x^{p_{ki}} \\
 \smash[t]{\stackrel{\eqref{eqn:IAC8:violate2}}{\Rightarrow}} x^{p_{jj}} &\nequiv d_{jk}d_{jj}^{-1}d_{ki}d_{kk}^{-1}x^{p_{ki}} \\
 \smash[t]{\stackrel{\eqref{eqn:IAC2}}{\Rightarrow}} x^{p_{jj}} &\nequiv d_{ii}d_{ij}^{-1}d_{ki}d_{kk}^{-1}d_{ij}d_{ii}^{-1}x^{p_{jj}} \\
\Rightarrow d_{ii}d_{jj}d_{kk} &\nequiv d_{ij}d_{jk}d_{ki} \Rightarrow \text{(ix)} \\
 \eqref{eqn:mac-X1}^\dagger:  d_{jk}x^{p_{jk}} &\nequiv d_{ji}x^{p_{ji}} \\
 \smash[t]{\stackrel{\eqref{eqn:IAC1}}{\Rightarrow}} x^{p_{jk}} &\nequiv d_{ji}d_{jk}^{-1}d_{ik}d_{ii}^{-1}x^{p_{jk}} \\
 \Rightarrow 0 &\nequiv \det(\boldsymbol{D}_{i,j,i,k}) \Rightarrow \text{(v)} \\
 \eqref{eqn:mac-X2}^\dagger: d_{jj}x^{p_{jj}} &\nequiv d_{jk}x^{p_{jk}} \\
 \smash[t]{\stackrel{\eqref{eqn:IAC8:violate2}}{\Rightarrow}} x^{p_{jj}} &\nequiv d_{jk}d_{jj}^{-1}d_{ki}d_{kk}^{-1}x^{p_{ki}} \\
 \smash[t]{\stackrel{\eqref{eqn:IAC2}}{\Rightarrow}} x^{p_{jj}} &\nequiv d_{jk}d_{jj}^{-1}d_{ki}d_{kk}^{-1}d_{ij}d_{ii}^{-1}x^{p_{jj}} \\
 \Rightarrow d_{ii}d_{jj}d_{kk} &\nequiv d_{ij}d_{jk}d_{ki} \Rightarrow \text{(ix)} \\
 \eqref{eqn:mac-X3}^\dagger: d_{jj}x^{p_{jj}} &\nequiv d_{ji}x^{p_{ji}} \\
 \smash[t]{\stackrel{\eqref{eqn:IAC2}}{\Rightarrow}} x^{p_{ji}} &\nequiv d_{jj}d_{ji}^{-1}d_{ii}d_{ij}^{-1}x^{p_{ki}} \\
 \smash[t]{\stackrel{\eqref{eqn:IAC3}}{\Rightarrow}} x^{p_{ji}} &\nequiv d_{jj}d_{ji}^{-1}d_{ii}d_{ij}^{-1}d_{jj}d_{ji}^{-1}x^{p_{ki}} \\
 \smash[t]{\stackrel{\eqref{eqn:IAC1}}{\Rightarrow}} x^{p_{ji}} &\nequiv d_{jj}d_{ji}^{-1}d_{ii}d_{ij}^{-1}d_{jj}d_{ji}^{-1}d_{ii}d_{ij}^{-1}x^{p_{ji}} \\
 \smash[t]{\stackrel{(ii)}{\Rightarrow}} d_{ii}d_{jj}d_{kk}d_{kk} &\nequiv d_{ik}d_{ki}d_{jk}d_{kj} \Rightarrow \text{(x)} \\
 \eqref{eqn:inter-X1}^\dagger: d_{jj} x^{p_{jj}} &\nequiv d_{jk}x^{p_{ik}} \\
 \smash[t]{\stackrel{\eqref{eqn:IAC2}}{\Rightarrow}} x^{p_{jk}} &\nequiv d_{jj}d_{jk}^{-1}d_{ii}d_{ij}^{-1}x^{p_{ki}} \\
 \smash[t]{\stackrel{\eqref{eqn:IAC8:violate2}}{\Rightarrow}} x^{p_{jk}} &\nequiv d_{jj}d_{jk}^{-1}d_{ii}d_{ij}^{-1}d_{kk}d_{ki}^{-1}x^{p_{jk}} \\
 d_{ii}d_{jj}d_{kk} &\nequiv d_{ij}d_{jk}d_{ki} \Rightarrow \text{(ix)} \\
 \eqref{eqn:inter-X2}^\dagger: d_{jj} x^{p_{jj}} &\nequiv d_{jk}x^{p_{kk}} \\
 \smash[t]{\stackrel{\eqref{eqn:IAC2}}{\Rightarrow}} x^{p_{jj}} &\nequiv d_{jk}d_{jj}^{-1}d_{ij}d_{ik}^{-1}x^{p_{jj}} \\
 \Rightarrow 0 &\nequiv \det(\boldsymbol{D}_{i,j,j,k}) \Rightarrow \text{(vi)} \\
 \eqref{eqn:inter-X3}^\dagger: d_{jj} x^{p_{jj}} &\nequiv d_{ji}x^{p_{ki}} \\
 \smash[t]{\stackrel{\eqref{eqn:IAC2}}{\Rightarrow}} x^{p_{jj}} &\nequiv d_{ji}d_{jj}^{-1}d_{ij}d_{ii}^{-1}x^{p_{jj}} \\
 \Rightarrow 0 &\nequiv \det(\boldsymbol{D}_{i,j,i,j}) \Rightarrow \text{(iv)} \\
 \eqref{eqn:inter-X4}^\dagger: d_{jj} x^{p_{jj}} &\nequiv d_{ji}x^{p_{ii}} \\
 \smash[t]{\stackrel{\eqref{eqn:IAC6}}{\Rightarrow}} x^{p_{jj}} &\nequiv d_{ji}d_{jj}^{-1}d_{kj}d_{ki}^{-1}x^{p_{jj}} \\
 \Rightarrow 0 &\nequiv \det(\boldsymbol{D}_{i,j,j,k}) \Rightarrow \text{(vi)} \\
 \eqref{eqn:inter-X5}^\dagger: d_{jk} x^{p_{jk}} &\nequiv d_{jj}x^{p_{kj}} \\
 \smash[t]{\stackrel{\eqref{eqn:IAC1}}{\Rightarrow}} x^{p_{jk}} &\nequiv d_{jj}d_{jk}^{-1}d_{ik}d_{ij}^{-1}x^{p_{jk}} \\
 \Rightarrow 0 &\nequiv \det(\boldsymbol{D}_{i,j,j,k}) \Rightarrow \text{(vi)} \\
 \eqref{eqn:inter-X6}^\dagger: d_{jk} x^{p_{jk}} &\nequiv d_{jj}x^{p_{ij}}
\end{align*}
$\eqref{eqn:inter-X6}^\dagger$ contradicts \eqref{eqn:IAC5:violate1} and is treated separately.
\begin{align*}
 \eqref{eqn:inter-X7}^\dagger: d_{jk} x^{p_{jk}} &\nequiv d_{ji}x^{p_{ki}} \\
 \smash[t]{\stackrel{\eqref{eqn:IAC8:violate2}}{\Rightarrow}} x^{p_{jk}} &\nequiv d_{ji}d_{jk}^{-1}d_{kk}d_{ki}^{-1}x^{p_{jk}} \\
 \Rightarrow 0 &\nequiv \det(\boldsymbol{D}_{k,j,k,i}) \Rightarrow \text{(vii)} \\
 \eqref{eqn:inter-X8}^\dagger: d_{jk} x^{p_{jk}} &\nequiv d_{ji}x^{p_{ii}} \\
 \smash[t]{\stackrel{\eqref{eqn:IAC6}}{\Rightarrow}} x^{p_{jk}} &\nequiv d_{ji}d_{jk}^{-1}d_{kj}d_{ki}^{-1}x^{p_{jj}} \\
 \smash[t]{\stackrel{\eqref{eqn:IAC2}}{\Rightarrow}} x^{p_{jk}} &\nequiv d_{ji}d_{jk}^{-1}d_{kj}d_{ki}^{-1}d_{ii}d_{ij}^{-1}x^{p_{ki}} \\
 \smash[t]{\stackrel{\eqref{eqn:IAC8:violate2}}{\Rightarrow}} x^{p_{jk}} &\nequiv d_{ji}d_{jk}^{-1}d_{kj}d_{ki}^{-1}d_{ii}d_{ij}^{-1}d_{kk}d_{ki}^{-1}x^{p_{jk}} \\
\smash[t]{\stackrel{(i)}{\Rightarrow}} d_{ji}d_{ik}d_{kj}d_{ki} &\nequiv d_{ii}d_{kk}d_{ji}d_{kj}
\end{align*}
\begin{align*}
\Rightarrow d_{ik}d_{ki} &\nequiv d_{ii}d_{kk} \\
\Rightarrow 0 &\nequiv \det(\boldsymbol{D}_{k,i,k,i}) \Rightarrow \text{(iii)}
\end{align*}
Analogously, we consider the cyclically relabelled versions $i \rightarrow k \rightarrow j \rightarrow i$ of the separability conditions \eqref{eqn:intra-X1}$^\star$ to \eqref{eqn:inter-X8}$^\star$, with superscript symbol $^\star$ denoting the relabelled versions.
\begin{align*}
 \eqref{eqn:intra-X1}^\star: x^{p_{ik}} & \nequiv x^{p_{jk}} \\
 \smash[t]{\stackrel{\eqref{eqn:IAC8:violate2}}{\Rightarrow}} x^{p_{ik}} &\nequiv d_{ki}d_{kk}^{-1}x^{p_{ki}} \\
 \smash[t]{\stackrel{\eqref{eqn:IAC3}}{\Rightarrow}} x^{p_{ik}} &\nequiv d_{ki}d_{kk}^{-1}d_{jk}d_{ji}^{-1}x^{p_{ik}} \\
 \Rightarrow 0 &\nequiv \det(\boldsymbol{D}_{k,j,k,i}) \Rightarrow \text{(vii)} \\
        \eqref{eqn:intra-X2}^\star: x^{p_{kk}} &\nequiv x^{p_{ik}} \\
 \smash[t]{\stackrel{\eqref{eqn:IAC3}}{\Rightarrow}} x^{p_{kk}} &\nequiv d_{ji}d_{jk}^{-1}x^{p_{ki}} \\
 \smash[t]{\stackrel{\eqref{eqn:IAC2}}{\Rightarrow}} x^{p_{kk}} &\nequiv d_{ji}d_{jk}^{-1}d_{ik}d_{ii}^{-1}x^{p_{kk}} \\
 \Rightarrow 0 &\nequiv \det(\boldsymbol{D}_{k,j,i,k}) \Rightarrow \text{(v)} \\
 \eqref{eqn:intra-X3}^\star: x^{p_{kk}} &\nequiv x^{p_{jk}} \\
 \smash[t]{\stackrel{\eqref{eqn:IAC8:violate2}}{\Rightarrow}} x^{p_{kk}} &\nequiv d_{ki}d_{kk}^{-1}x^{p_{ki}} \\
 \smash[t]{\stackrel{\eqref{eqn:IAC2}}{\Rightarrow}} x^{p_{kk}} &\nequiv d_{ki}d_{kk}^{-1}d_{ik}d_{ii}^{-1}x^{p_{kk}} \\
 \Rightarrow 0 &\nequiv \det(\boldsymbol{D}_{i,k,i,k}) \Rightarrow \text{(iii)} \\
 \eqref{eqn:mac-X1}^\star:  d_{ki}x^{p_{ki}} &\nequiv d_{kj}x^{p_{kj}} \\
 \smash[t]{\stackrel{\eqref{eqn:IAC3}}{\Rightarrow}} x^{p_{kk}} &\nequiv d_{kj}d_{ki}^{-1}d_{ji}d_{jj}^{-1}x^{p_{ki}} \\
 \Rightarrow 0 &\nequiv \det(\boldsymbol{D}_{i,j,j,k}) \Rightarrow \text{(vi)} \\
 \eqref{eqn:mac-X2}^\star: d_{kk}x^{p_{kk}} &\nequiv d_{ki}x^{p_{ki}} \\
\smash[t]{\stackrel{\eqref{eqn:IAC2}}{\Rightarrow}} x^{p_{kk}} &\nequiv d_{ki}d_{kk}^{-1}d_{ik}d_{ii}^{-1}x^{p_{kk}} \\
 \Rightarrow 0 &\nequiv \det(\boldsymbol{D}_{i,k,i,k}) \Rightarrow \text{(iii)} \\
 \eqref{eqn:mac-X3}^\star: d_{kk}x^{p_{kk}} &\nequiv d_{kj}x^{p_{kj}} \\
 \smash[t]{\stackrel{\eqref{eqn:IAC3}}{\Rightarrow}} x^{p_{kk}} &\nequiv d_{kj}d_{kk}^{-1}d_{ji}d_{jj}^{-1}x^{p_{ki}} \\
 \smash[t]{\stackrel{\eqref{eqn:IAC2}}{\Rightarrow}} x^{p_{kk}} &\nequiv d_{kj}d_{kk}^{-1}d_{ji}d_{jj}^{-1}d_{ik}d_{ii}^{-1}x^{p_{kk}} \\
 \Rightarrow d_{ii}d_{jj}d_{kk} &\nequiv d_{ji}d_{kj}d_{ik} \Rightarrow \text{(ix)} \\
 \eqref{eqn:inter-X1}^\star: d_{kk} x^{p_{kk}} &\nequiv d_{ki}x^{p_{ji}} \\
 \smash[t]{\stackrel{\eqref{eqn:IAC1}}{\Rightarrow}} x^{p_{kk}} &\nequiv d_{ki}d_{kk}^{-1}d_{ij}d_{ii}^{-1}x^{p_{kj}} \\ 
 \smash[t]{\stackrel{\eqref{eqn:IAC3}}{\Rightarrow}} x^{p_{kk}} &\nequiv d_{ki}d_{kk}^{-1}d_{ij}d_{ii}^{-1}d_{ji}d_{jj}^{-1}x^{p_{ki}} \\
 \smash[t]{\stackrel{\eqref{eqn:IAC2}}{\Rightarrow}} x^{p_{kk}} &\nequiv  d_{ki}d_{kk}^{-1}d_{ij}d_{ii}^{-1}d_{ji}d_{jj}^{-1}d_{ik}d_{ii}^{-1}x^{p_{kk}} \\ 
 \Rightarrow d_{ii}d_{jj}d_{kk}d_{ii} &\nequiv d_{ki}d_{ij}d_{ji}d_{ik} \Rightarrow \text{(x)} \\
 \eqref{eqn:inter-X2}^\star: d_{kk} x^{p_{kk}} &\nequiv d_{ki}x^{p_{ii}} \\
 \smash[t]{\stackrel{\eqref{eqn:IAC4}}{\Rightarrow}} x^{p_{kk}} &\nequiv d_{ki}d_{kk}^{-1}d_{jk}d_{ji}^{-1}x^{p_{kk}} \\
 \Rightarrow 0 &\nequiv \det(\boldsymbol{D}_{k,j,k,i}) \Rightarrow \text{(vii)} \\
 \eqref{eqn:inter-X3}^\star: d_{kk} x^{p_{kk}} &\nequiv d_{kj}x^{p_{ij}} \\
 \smash[t]{\stackrel{\eqref{eqn:IAC5:violate1}}{\Rightarrow}} x^{p_{kk}} &\nequiv d_{kj}d_{kk}^{-1}d_{jk}d_{jj}^{-1}x^{p_{jk}} \\
 \smash[t]{\stackrel{\eqref{eqn:IAC8:violate2}}{\Rightarrow}} x^{p_{kk}} &\nequiv d_{kj}d_{kk}^{-1}d_{jk}d_{jj}^{-1}d_{ki}d_{kk}^{-1}x^{p_{jk}} \\
 \smash[t]{\stackrel{\eqref{eqn:IAC2}}{\Rightarrow}} x^{p_{kk}} &\nequiv d_{kj}d_{kk}^{-1}d_{jk}d_{jj}^{-1}d_{ki}d_{kk}^{-1}d_{ik}d_{ii}^{-1}x^{p_{jk}} \\
 \Rightarrow d_{ii}d_{jj}d_{kk}d_{kk} &\nequiv d_{kj}d_{jk}d_{ki}d_{ik} \Rightarrow \text{(x)} \\
 \eqref{eqn:inter-X4}^\star: d_{kk} x^{p_{kk}} &\nequiv d_{kj}x^{p_{jj}} \\
 \smash[t]{\stackrel{\eqref{eqn:IAC2}}{\Rightarrow}} x^{p_{kk}} &\nequiv d_{kj}d_{kk}^{-1}d_{ik}d_{ij}^{-1}x^{p_{kk}} \\
 \Rightarrow 0 &\nequiv \det(\boldsymbol{D}_{k,j,k,i}) \Rightarrow \text{(vii)}
\end{align*}
\begin{align*}
 \eqref{eqn:inter-X5}^\star: d_{ki} x^{p_{ki}} &\nequiv d_{kk}x^{p_{ik}} \\
 \smash[t]{\stackrel{\eqref{eqn:IAC8:violate2}}{\Rightarrow}} x^{p_{ki}} &\nequiv d_{kk}d_{ki}^{-1}d_{ji}d_{jk}^{-1}x^{p_{ki}} \\
 \Rightarrow 0 &\nequiv \det(\boldsymbol{D}_{k,j,k,i}) \Rightarrow \text{(vii)} \\
 \eqref{eqn:inter-X6}^\star: d_{ki} x^{p_{ki}} & \nequiv d_{kk}x^{p_{jk}}
\end{align*}
\eqref{eqn:inter-X6}$^\star$ contradicts \eqref{eqn:IAC8:violate2} and is treated separately.
\begin{align*}
 \eqref{eqn:inter-X7}^\star: d_{ki} x^{p_{ki}} &\nequiv d_{kj}x^{p_{ij}} \\
 \smash[t]{\stackrel{\eqref{eqn:IAC5:violate1}}{\Rightarrow}} x^{p_{ki}} &\nequiv d_{kj}d_{ki}^{-1}d_{jk}d_{jj}^{-1}x^{p_{jk}} \\
 \smash[t]{\stackrel{\eqref{eqn:IAC8:violate2}}{\Rightarrow}} x^{p_{ki}} &\nequiv d_{kj}d_{ki}^{-1}d_{jk}d_{jj}^{-1}d_{ki}d_{kk}^{-1}x^{p_{jk}} \\
 \Rightarrow 0 &\nequiv \det(\boldsymbol{D}_{k,j,k,j}) \Rightarrow \text{(viii)} \\
 \eqref{eqn:inter-X8}^\star: d_{ki} x^{p_{ki}} &\nequiv d_{kj}x^{p_{jj}} \\
 \smash[t]{\stackrel{\eqref{eqn:IAC2}}{\Rightarrow}} x^{p_{ki}} &\nequiv d_{kj}d_{ki}^{-1}d_{ii}d_{ij}^{-1}x^{p_{ki}} \\
 \Rightarrow 0 &\nequiv \det(\boldsymbol{D}_{i,j,i,k}) \Rightarrow \text{(v)}
\end{align*}
The constraints (i) and (ii) are proven as follows.
\begin{align*}
 \eqref{eqn:IAC2}: d_{ij}x^{p_{jj}} &\equiv d_{ik}x^{p_{kk}}  \\
 \smash[t]{\stackrel{\eqref{eqn:IAC3}}{\Rightarrow}} d_{ij}x^{p_{jj}} &\equiv d_{ik}d_{ji}d_{jk}^{-1}x^{p_{ii}}  \\
 \smash[t]{\stackrel{\eqref{eqn:IAC6}}{\Rightarrow}} d_{ij}x^{p_{jj}} &\equiv d_{ik}d_{ji}d_{jk}^{-1}d_{kj}d_{ki}^{-1}x^{p_{jj}}  \\
 \Rightarrow d_{ij}d_{jk}d_{ki} &\equiv d_{ik}d_{kj}d_{ji} \Rightarrow \text{(i)} \\
 \eqref{eqn:IAC1}: d_{ik}x^{p_{jk}} &\equiv d_{ii}x^{p_{ji}} \\
 \smash[t]{\stackrel{\eqref{eqn:IAC5:violate1}}{\Rightarrow}} d_{ik}d_{jj}d_{jk}^{-1}x^{p_{ij}} &\equiv d_{ii}x^{p_{ji}} \\
 \smash[t]{\stackrel{\eqref{eqn:IAC4}}{\Rightarrow}} d_{ik}d_{jj}d_{jk}^{-1}x^{p_{ij}} &\equiv d_{ii}d_{kj}d_{ki}^{-1}x^{p_{ij}} \\
 \Rightarrow d_{jj}d_{ik}d_{ki} &\equiv d_{ii}d_{jk}d_{kj} \Rightarrow \text{(ii)} \\
 \eqref{eqn:IAC2}: d_{ii}x^{p_{ki}} &\equiv d_{ij}x^{p_{jj}} \\
 \smash[t]{\stackrel{\eqref{eqn:IAC6}}{\Rightarrow}} d_{ii}x^{p_{ki}} &\equiv d_{ij}d_{kk}d_{kj}^{-1}x^{p_{ik}} \\
 \smash[t]{\stackrel{\eqref{eqn:IAC3}}{\Rightarrow}} d_{ii}x^{p_{ki}} &\equiv d_{ij}d_{kk}d_{kj}^{-1}d_{ji}d_{jk}^{-1}x^{p_{ki}} \\
 \Rightarrow d_{ii}d_{jk}d_{kj} &\equiv d_{kk}d_{ij}d_{ji} \Rightarrow \text{(ii)} \\
 \eqref{eqn:IAC3}: d_{jj}x^{p_{kj}} &\equiv d_{ji}x^{p_{ki}} \\
 \smash[t]{\stackrel{\eqref{eqn:IAC8:violate2}}{\Rightarrow}} d_{jj}x^{p_{kj}} &\equiv d_{ji}d_{kk}d_{ki}^{-1}x^{p_{jk}} \\
 \smash[t]{\stackrel{\eqref{eqn:IAC1}}{\Rightarrow}} d_{jj}x^{p_{kj}} &\equiv d_{ji}d_{kk}d_{kj}^{-1}d_{ij}d_{ik}^{-1}x^{p_{kj}} \\
 \Rightarrow d_{ii}d_{jk}d_{kj} &\equiv d_{kk}d_{ij}d_{ji} \Rightarrow \text{(ii)}
\end{align*}
\end{appendix}

\bibliographystyle{IEEEtran}
\bibliography{paper}

\end{document}